\documentclass[10pt]{article}\usepackage[]{graphicx}\usepackage[]{xcolor}
\makeatletter
\def\maxwidth{ %
  \ifdim\Gin@nat@width>\linewidth
    \linewidth
  \else
    \Gin@nat@width
  \fi
}
\makeatother

\definecolor{fgcolor}{rgb}{0.345, 0.345, 0.345}

\usepackage{framed}
\makeatletter
 {\par\unskip\endMakeFramed%
 \at@end@of@kframe}
\makeatother

\definecolor{shadecolor}{rgb}{.97, .97, .97}
\definecolor{messagecolor}{rgb}{0, 0, 0}
\definecolor{warningcolor}{rgb}{1, 0, 1}
\definecolor{errorcolor}{rgb}{1, 0, 0}
\newenvironment{knitrout}{}{} 

\usepackage{alltt}
\usepackage{color}
\usepackage{graphicx} 
\usepackage{booktabs} 
\usepackage{multirow}
\usepackage{makecell}
\usepackage{array} 
\usepackage{subcaption}
\usepackage{setspace} 
\usepackage{float} 
\usepackage{tabularx}
\usepackage[round]{natbib}
\usepackage[font={small,it}]{caption}
\usepackage[utf8]{inputenc}
\usepackage{mathtools}
\usepackage{fullpage}
\usepackage{verbatim}
\usepackage{amsfonts,amssymb,amsthm}
\usepackage{url}
\usepackage[colorlinks=true,citecolor=blue]{hyperref}

\IfFileExists{upquote.sty}{\usepackage{upquote}}{}
\begin{document}

\title{Poisson approximate likelihood compared to the particle filter}
\author{Yize Hao, Aaron A. Abkemeier and Edward L. Ionides}
\date{The University of Michigan}
\maketitle

\begin{abstract}
Filtering algorithms are fundamental for inference on partially observed stochastic dynamic systems, since they provide access to the likelihood function and hence enable likelihood-based or Bayesian inference.
A novel Poisson approximate likelihood (PAL) filter was introduced by \cite{wwr}.
PAL employs a Poisson approximation to conditional densities, offering a fast approximation to the likelihood function for a certain subset of partially observed Markov process models.
A central piece of evidence for PAL is the comparison in Table~1 of \cite{wwr}, which claims a large improvement for PAL over a standard particle filter algorithm.
This evidence, based on a model and data from a previous scientific study by \citet{stocks}, might suggest that researchers confronted with similar models should use PAL rather than particle filter methods.
Taken at face value, this evidence also reduces the credibility of \cite{stocks} by indicating a shortcoming with the numerical methods that they used.
However, we show that the comparison of log-likelihood values made by \cite{wwr} is flawed because their PAL calculations were carried out using a dataset scaled differently from the previous study.
If PAL and the particle filter are applied to the same data, the advantage claimed for PAL disappears.
On simulations where the model is correctly specified, the particle filter outperforms PAL.

\end{abstract}













This article results from an investigation of the results presented by \citet{wwr} (henceforth, WWR) in their Table~1.
WWR were given the opportunity to submit a correction, after we shared the results of our investigation with them, but they declined.
The theory developed by WWR shows that their Poisson Approximate Likelihood (PAL) method has some potentially useful scaling properties.
This theory is supported by numerical results, in their Table~1, which erroneously claim to show that PAL has substantially stronger performance than a particle filter (PF) on an example of scientific interest.
We present corrected results so that researchers considering whether to implement PAL are appropriately informed about its benefits.

Table~1 of WWR uses a model and data adapted from \citet{stocks} (henceforth, SBH).
SBH used PF to calculate the likelihood for a stochastic dynamic model of rotavirus transmission.
SBH found clear evidence for the importance of including overdispersion in the model for their epidemiological data.
This is significant because most earlier research on population dynamics avoided consideration of overdispersion, perhaps due to the lack of available statistical methodology for fitting overdispersed nonlinear stochastic dynamic models. 
The conclusions of SBH hinge on a comparison of likelihoods, and so the results of WWR discredit those conclusions by indicating that SBH based their reasoning on inaccurately computed likelihoods.
An important consequence of correcting Table~1 of WWR is that the results of SBH stand undiminished. 

SBH and WWR each fitted three different rotavirus models.
The first has equidispersion (i.e., no overdispersion) in the measurement model and the dynamic model, and is called EqEq by WWR.
The second, EqOv, includes overdispersion in only the measurement model.
The third, OvOv, includes overdispersion in both these model components.
We focus on OvOv, which WWR and SBH both found to be the best fitting model. 

We show that the claimed advantage for PAL over PF, on the OvOv model, arose because WWR used a different scaling of the data from SBH.
Two models for the same data can properly be compared by their likelihood, even if the models have entirely different structures.
Allowance for the number of estimated parameters can be made using a quantity such as Akaike's information criterion \citep{Akaike1974ANL}.
However, if data are rescaled, an adjustment is required to make likelihoods comparable.
For example, if one model describes a dataset in grams and another describes it in kilograms, then the latter model will earn an increased log-likelihood of $\log(10^3)$ for each data point simply because of the change in scale.
Presenting a direct comparison of a likelihood for the data in grams with a likelihood for the data in kilograms would evidently be inappropriate.


\begin{table}[ht] 
\centering 
\caption{AIC for the OvOv rotavirus model, computed using two filtering methods. PAL is the Poisson approximate likelihood, implemented using the code of WWR. PF is the particle filter, implemented using the R package pomp \citep{pomppackagepaper}. Lines 1, 2 and 7 are taken from WWR, and the remainder are our own computations. Line 3 recomputes the previously published value in Line 2, and the small difference is presumably due to rounding in Table~2 of WWR. We used $5 \times 10^4$ particles for both PF and PAL. PF was repeated 36 times to reduce the Monte Carlo variance, but this step was not necessary for PAL due to its lower Monte~Carlo variance. PF results were maximized using iterated filtering, following the approach of SBH. PAL results were maximized using coordinate gradient descent, using the code of WWR.
} 
\label{tab:ovovrealdata}
\begin{tabular}{lllllr} 
  \hline
  & Method & Data & Model & Parameters                          & AIC  \\
  \hline
  1. & PF  & Rescaled counts  & SBH   & Table~2 of SBH &  20134
  \\
  2. & PAL & Counts  & WWR   & Table~2 of WWR                      & 13778
  \\
  3. & PAL & Counts  & WWR   & Table~2 of WWR  &
    13799 
  \\
  4. & PF  & Counts & WWR    & Table~2 of WWR &
    14549
  \\
  5. & PF  & Counts & WWR modified   & Maximum likelihood &
     13768
  \\
  6. & PAL & Counts & WWR modified   & Matching line 5 &
     13937
  \\   
  7. & Benchmark & Rescaled counts & log-ARMA(2,1)    &  & 23043
  \\
  8. & Benchmark & Counts & log-ARMA(2,1)      &  &
    12751 
  \\
 \hline
\end{tabular}
\end{table}

SBH fitted their model to a dataset of rescaled counts derived by dividing the original reported count data by an estimated reporting rate.
Thus, SBH fitted to data on the scale of the disease incidence in the population.
By contrast, WWR fitted directly to the reported case count data.
The reporting rate used by SBH varied over time and location, but was generally around $7\%$.
On $3\times 416$ data points, this corresponds to a discrepancy of $-1248 \, \log(0.07) \approx 3300$ log-likelihood units, largely explaining the difference interpreted by WWR as evidence supporting PAL (Table~\ref{tab:ovovrealdata}, comparing lines 1 and 2).
The comparison between PAL and PF can be corrected by applying the method of SBH to the model and data of WWR, or vice versa.
Since the method of SBH is applicable to a more general class of models, and supported by widely-used software, it was convenient to apply the SBH method to the model and data of WWR.
The large discrepancy in log-likelihood disappears when recomputing the likelihood using PF for the model fitted via PAL (Table~\ref{tab:ovovrealdata}, comparing lines 3 and 4).
Some discrepancy remains, and we continued our investigation to establish the cause of this.

Inspection of log-likelihood anomalies \citep{wheeler24,li24} showed that the initial conditions for the latent process in January 2001 were fixed at values which were incompatible with the trajectory of the data early in the time series \citep[Figure~3 of][]{hao24}.
By contrast, SBH fixed their initial conditions 6 years before the first measurement, giving time for the system to reach its equilibrium distribution.
Line 5 of Table~\ref{tab:ovovrealdata} incorporates the SBH specification of initial values into the model of WWR.
The likelihood for this modified model was then maximized using an iterated filtering procedure similar to SBH.
Comparison of lines 3 and 5 shows that this improvement enables PF to reach the AIC values attained by PAL, and show a small improvement.
Further, at these maximum likelihood parameter values, we found that PF beats PAL (comparing lines 5 and 6; for line 6 we initialized PAL at the average value of the stochastic initialization used in line 5).

WWR compared their fitted models with a log-ARMA(2,1) benchmark AIC value of 23043 (line 7).
From this high AIC value, they inferred that all the mechanistic models possessing overdispersion have better statistical fit than a simple log-linear time series model.
However, this AIC value corresponds to SBH's data, not the data fitted by WWR.
Thus, line 7 can be properly compared only to line 1 but not to any other line of  Table~\ref{tab:ovovrealdata}. 
We refitted a log-ARMA(2,1) model to the SBH data and obtained a similar AIC value (23085).
Carrying out the same computation for the data fitted by WWR gives a log-ARMA benchmark AIC value of 12751 (line 8).
Thus, the statistical fit of the mechanistic model considered by WWR is inferior to a simple log-ARMA model.
This holds for all the variants in lines 2--6 of  Table~\ref{tab:ovovrealdata}, regardless of whether PF or PAL is used.
The goal of mechanistic modeling is not necessarily to beat a simple statistical benchmark, but falling far below a simple statistical benchmark is an indication that additional model development could be worthwhile \citep{wheeler24}.
A correct interpretation of Table~\ref{tab:ovovrealdata} is therefore very different to the conclusions drawn by WWR, who compared line 2 inappropriately to lines 1 and 7.

\begin{knitrout}
\definecolor{shadecolor}{rgb}{0.969, 0.969, 0.969}\color{fgcolor}\begin{figure}

{\centering \includegraphics[width=4.5in]{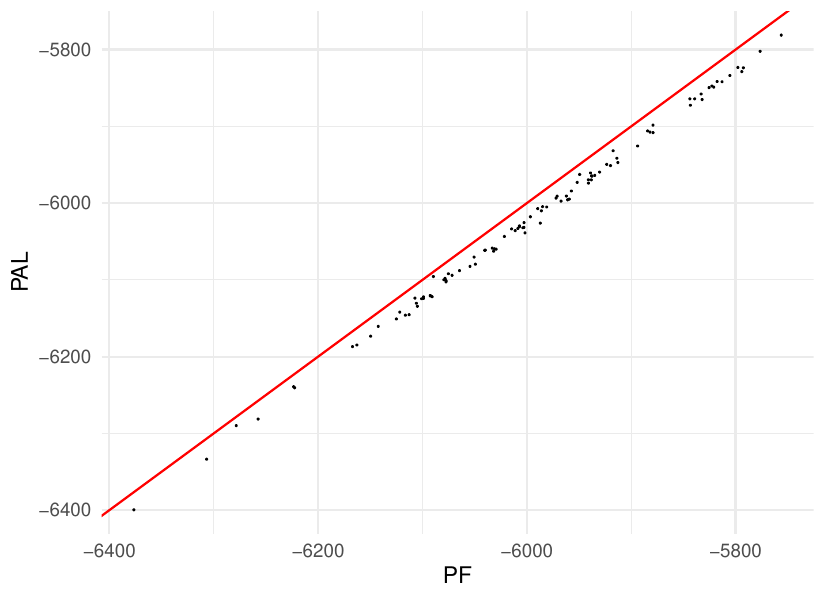} 

}

\caption[Log-likelihood computed using PAL and PF for 100 randomly simulated datasets using the model and parameter values of WWR]{Log-likelihood computed using PAL and PF for 100 randomly simulated datasets using the model and parameter values of WWR. Simulations with one or more zero counts were disqualified since they resulted in errors for the PAL implementation of WWR. We used $5 \times 10^4$ particles for both PF and PAL. PF was repeated 36 times to reduce the Monte Carlo variance, but this step was not necessary for PAL due to its lower Monte~Carlo variance. The red line corresponds to equality of the two estimates.}\label{fig:plot1-100-sim-compare}
\end{figure}

\end{knitrout}

In the presence of model misspecification, it becomes difficult to compare likelihood evaluation methods.
A likelihood approximation, such as PAL, may potentially obtain a higher value than the exact likelihood if it compensates for model misspecification.
Log-likelihood is a proper scoring rule for forecasts \citep{gneiting2}, and both the particle filter and PAL construct their log-likelihood estimates via a sequence of one-step forecasts.
Therefore, if the model is correctly specified, the approximation error in PAL can only decrease the expected log-likelihood.
We tested this on simulated data for which the model of WWR is correctly specified.
For this simulation study, the particle filter out-performs PAL (Figure~\ref{fig:plot1-100-sim-compare}).
On average, the particle filter likelihood estimate is $25.1$ log units higher than the PAL.
We know from the benchmark AIC value in Table~\ref{tab:ovovrealdata} (comparing line 8 to lines 2--6) that there is substantial model misspecification.


It is currently unclear why the model of SBH beats the log-ARMA benchmark for their data, whereas the model of WWR fits more poorly than the log-ARMA model for its corresponding data.
Additional rounds of model development are required to resolve this, for which it is desirable to employ statistical methods that are broadly applicable both in theory and practice.
Particle filter methods meet this criterion since they have the plug-and-play property \citep{breto09,he10} which is not possessed by PAL.
Although WWR have shown that PAL is a potentially useful algorithm with some favorable theoretical properties, the corrected evidence does not indicate an advantage for using PAL in situations where the particle filter is effective.

The source code for this article is available on GitHub (\url{https://github.com/ionides/pal-vs-pf}) and archived at Zenodo (\url{https://zenodo.org/doi/10.5281/zenodo.13777112}).
An extended description of our methods, together with additional numerical results, is provided by \cite{hao24}.

\bibliography{bib-pal}

\clearpage

\end{document}